\documentclass[aps,prb,reprint]{revtex4-2}

\usepackage{graphicx}
\usepackage{bm}
\usepackage{bbold}
\usepackage{amssymb}
\usepackage{amsmath}

\usepackage{color}

\renewcommand{\vec}[1]{{\mathbf #1}}

\bibpunct{[}{]}{,}{n}{}{}

\begin{document}

\title{Higher-order localization landscape theory of Anderson localization}

\author{Sergey E. Skipetrov}
\email[]{Sergey.Skipetrov@lpmmc.cnrs.fr}
\affiliation{Univ. Grenoble Alpes, CNRS, LPMMC, 38000 Grenoble, France}

\date{\today}

\begin{abstract}
For a Hamiltonian ${\hat H}$ containing a position-dependent (disordered) potential, we introduce a sequence of landscape functions $u_n(\vec{r})$ obeying ${\hat H} u_n(\vec{r}) = u_{n-1}(\vec{r})$ with $u_0(\vec{r}) = 1$. For $n \to \infty$, $1/v_n(\vec{r}) = u_{n-1}(\vec{r})/u_{n}(\vec{r})$ converges to the lowest eigenenergy $E_1$ of ${\hat H}$ whereas $u_{\infty}(\vec{r})$ yields the corresponding wave function $\psi_1(\vec{r})$. For large but finite $n$, $v_n(\vec{r})$ can be approximated by a piecewise constant function $v_n(\vec{r}) \simeq v_n^{(m)}$ for $\vec{r} \in \Omega_m$ and yields progressively improving estimations of eigenenergies $E_m = 1/v_n^{(m)}$ of locally fundamental eigenstates $\psi_m(\vec{r}) \propto u_{n}(\vec{r})$ in spatial domains $\Omega_m$. These general results are illustrated by a number of examples in one dimension: box potential, sequence of randomly placed infinite potential barriers, smooth and spatially uncorrelated random  potentials, quasiperiodic potential,
as well as for the uncorrelated random potential in two dimensions.
\end{abstract}

\maketitle

\vspace{-1cm}

\section{Introduction}
\label{form}

The nontrivial impact of disorder on the properties of quantum and, more generally, wave systems attracts the attention of physicists since the seminal work of Anderson \cite{anderson58}. Disorder-induced localized states, their properties and impact on measurable quantities have been subjects of numerous experimental studies in solid-state physics  \cite{lee85,kramer93},
electromagnetics and optics \cite{chabanov00,schwartz07,lahini08,riboli11,segev13},
acoustics \cite{hu08,cobus18}, and cold-atom physics \cite{chabe08,billy08,roati08,kondov11,jendr12,semeghini15}. On the theory side, scaling theory \cite{abrahams79} provides some qualitative understanding of Anderson localization but going beyond it requires either sophisticated analytical approaches for calculating statistical properties of quantities of interest \cite{lifshits88,efetov96,mirlin00,evers08,wolfle10} or resource-consuming numerical techniques for solving the problem for a given realization of disorder and then collecting statistics over many realizations (Monte-Carlo method) \cite{markos06,rodriguez10,pinski12,delande14,slevin14,carnio19,haberko20,yamilov23}.

An original ``localization landscape'' theory proposed by Filoche and Mayboroda \cite{filoche12} allows for gaining a new insight into the problem and reducing its computational complexity at the same time. The main idea of this theory is easy to understand. Consider the time-independent Schr\"{o}dinger equation with a potential $V(\vec{r})$ (in dimensionless form) \cite{landau}:
\begin{eqnarray}
{\hat H} \psi(\vec{r}) = \left[ -\Delta + V(\vec{r}) \right] \psi(\vec{r})  = E \psi(\vec{r})
\label{sch}
\end{eqnarray}
on a spatial domain $\Omega$ of dimenisonality $d$ with a boundary condition
\begin{eqnarray}
\left. \psi(\vec{r}) \right|_{\vec{r} \in \partial \Omega} = 0
\label{bc}
\end{eqnarray}
at the $(d-1)$-dimensional boundary $\partial \Omega$ of  $\Omega$. Localization landscape theory proposes to represent eigenfunctions of Eq.\ (\ref{sch}) as $\psi(\vec{r}) = u_1(\vec{r}) \phi_1(\vec{r})$ with the equation for $u_1(\vec{r})$:
\begin{eqnarray}
{\hat H} u_1(\vec{r})  = 1
\label{land}
\end{eqnarray}
being imposed and to be solved with the boundary condition (\ref{bc}). Equation for $\phi_1(\vec{r})$ follows from Eqs.\ (\ref{sch}) and (\ref{land}) \cite{arnold16}:
\begin{eqnarray}
&&-\frac{1}{u_1(\vec{r})^2} \bm{\nabla} \cdot
\left[u_1(\vec{r})^2 \bm{\nabla} \phi_1(\vec{r}) \right]
+ \frac{ \phi_1(\vec{r})}{u_1(\vec{r})} = E \phi_1(\vec{r})\;\;
\label{eqphi}
\end{eqnarray}
Notice that we have equipped the landscape function $u$ with a subscript `1' of which the meaning will be clear from the following.
 
We note that the transformation of Eq.\ (\ref{sch}) into Eqs.\ (\ref{land}) and (\ref{eqphi}) is exact and does not rely on any approximation. Equation (\ref{eqphi}) can be seen as analogous to Eq.\ (\ref{sch}) with a modified differential operator (or ``kinetic energy'') and a new effective potential $1/u_1(\vec{r})$ replacing $V(\vec{r})$. The key idea of localization landscape theory is to neglect  the kinetic energy term 
$-\bm{\nabla} \cdot[u_1^2 \bm{\nabla} \phi_1]/u_1^2$
in Eq.\ (\ref{eqphi}) and study properties of the solutions of Eq.\ (\ref{sch}) based on the potential landscape $1/u_1(\vec{r})$ only. This introduces an enormous simplification of the initial problem and strongly reduces its computational complexity because the differential equation (\ref{land}) is generally easier to solve than the eigenvalue problem (\ref{sch}). Successful applications of localization landscape theory include studies of localization effects in semiconductors \cite{filoche17,piccardo17,li17} and in ensembles of ultracold atoms placed in disordered optical potentials \cite{pelletier22,stellin23} as the most prominent examples. The theory has been extended to deal with many-body problems \cite{bala20} and provides an alternative way to understand the localization phase diagram of Anderson model in three dimensions \cite{filoche23}.
Alternative and more general ways of defining landscape potentials have been proposed \cite{steinerberger17,herviou20,steinerberger21}.

In the present paper we propose an extension of localization landscape theory that brings some of its predictions closer to exact results and opens new opportunities for studying the properties of solutions of Eq.\ (\ref{sch}), still without solving the full eigenvalue problem. Our main idea is to keep the kinetic energy term in Eq.\ (\ref{eqphi}) and to apply to this equation the same manipulation as the one applied to Eq.\ (\ref{sch}) in the original localization landscape theory. Namely, in Sec.\ \ref{iter} we represent $\phi_1$ as a product of a new, second-order landscape function and a new wave function $\phi_2$ for which a new equation is derived. This procedure is repeated iteratively resulting in a sequence of higher-order landscape functions $u_n(\vec{r})$. In the limit $n \to \infty$, iterations converge to exact results for the lowest eigenvalue $E_1$ of Eq.\ (\ref{sch}) and the corresponding eigenfunction $\psi_1(\vec{r})$ because they actually correspond to the inverse power method for finding the smallest eigenvalue of a matrix which, in our case, represents the operator ${\hat H}$ in the configuration space (see, e.g., Ref.\ \cite{bronson14}). What is less trivial is that results for large but finite $n$ provide progressively improving approximations for low-order $E_m$ and $\psi_m(\vec{r})$, which is illustrated by a number of examples in Sec.\ \ref{examples}.

\section{Higher-order landscapes by iteration}
\label{iter}

To introduce higher-order landscape potentials, we propose to apply to Eq.\ (\ref{eqphi}) the manipulations transforming Eq.\ (\ref{sch}) into Eqs.\ (\ref{land}) and (\ref{eqphi}). We represent $\phi_1(\vec{r}) = v_2(\vec{r}) \phi_2(\vec{r})$ leading to $\psi(\vec{r}) = u_1(\vec{r}) \phi_1(\vec{r}) = u_1(\vec{r}) v_2(\vec{r}) \phi_2(\vec{r}) = u_2(\vec{r}) \phi_2(\vec{r})$ with $u_2(\vec{r}) = u_1(\vec{r}) v_2(\vec{r})$.  Manipulations similar to those transforming Eq.\ (\ref{sch}) into Eqs.\ (\ref{land}) and (\ref{eqphi}) yield
\begin{eqnarray}
&&-\frac{1}{u_1^2} \bm{\nabla} \cdot
\left[u_1^2 \bm{\nabla} v_2 \right]
+ \frac{1}{u_1} v_2 = 1
\label{land1}
\\
&&-\frac{1}{(u_1 v_2)^2} \bm{\nabla} \cdot
\left[(u_1 v_2)^2 \bm{\nabla} \phi_2 \right]
+ \frac{1}{v_2} \phi_2 = E \phi_2
\label{eqtheta}
\end{eqnarray}
A conjecture is that neglecting the kinetic energy term in the latter equation and using the new landscape function $u_2(\vec{r}) = u_1(\vec{r}) v_2(\vec{r})$ instead of $u_1(\vec{r})$ may be a better approximation than the one in the standard localization landscape theory.

We do not have to stop at Eqs.\ (\ref{land1}) and (\ref{eqtheta}), and can iterate the procedure again and again by representing $\phi_2(\vec{r})$ as a product of a new $v_3(\vec{r})$  and a new wave function $\phi_3(\vec{r})$, and so on. At each stage of such iterations, a new landscape function $u_n$ is obtained from the previous $u_{n-1}$ and an equation for $\phi_n(\vec{r})$ is derived. This leads to the following equation for $v_n(\vec{r})$ and $\phi_n(\vec{r})$, $n \geq 1$:
\begin{eqnarray}
&&-\left(  \prod\limits_{i=1}^{n-1} \frac{1}{v_i^2} \right) \bm{\nabla} \cdot
\left[ \left(  \prod\limits_{i=1}^{n-1} v_i^2 \right) \bm{\nabla} v_n \right]
+ \frac{v_n}{v_{n-1}} = 1\;\;\;\;\;\;
\label{landn}
\\
&&-\left(  \prod\limits_{i=1}^{n} \frac{1}{v_i^2} \right) \bm{\nabla} \cdot
\left[ \left(  \prod\limits_{i=1}^{n} v_i^2 \right) \bm{\nabla} \phi_n\right]
+ \frac{\phi_n}{v_n}= E \phi_n
\label{phin}
\end{eqnarray}
 with $v_0(\vec{r}) = 1/V(\vec{r})$ and $v_1(\vec{r}) = u_1(\vec{r})$  . 

Equations (\ref{landn}) and (\ref{phin}) can be rewritten in terms of the landscape function of order $n$
\begin{eqnarray}
u_n(\vec{r}) &=& \prod\limits_{i = 1}^{n} v_i(\vec{r}), \;\; n \geq 1
\label{vn}
\end{eqnarray}
so that
\begin{eqnarray}
v_n(\vec{r}) = \frac{u_n(\vec{r})}{u_{n-1}(\vec{r})}
\label{un}
\end{eqnarray}
We obtain
\begin{eqnarray}
&&{\hat H} u_n(\vec{r}) = u_{n-1}(\vec{r}) \text{~or~}
{\hat H}^n u_n(\vec{r}) = 1,\,\, n \geq 1
\label{equnfinal}
\\
&&-\frac{1}{u_n(\vec{r})^2} \bm{\nabla} \cdot
\left[ u_n(\vec{r})^2 \bm{\nabla} \phi_n(\vec{r}) \right]
+ \frac{u_{n-1}(\vec{r})}{u_n(\vec{r})} \phi_n(\vec{r})
\nonumber \\
&&\hphantom{-\frac{1}{u_n(\vec{r})^2} \bm{\nabla} \cdot
\left[ u_n(\vec{r})^2 \bm{\nabla} \phi_n(\vec{r}) \right]}
 = E \phi_n(\vec{r})
\label{eqphinfinal}
\end{eqnarray}
with $u_0(\vec{r}) = 1$ and a boundary condition 
\begin{eqnarray}
\left. u_n(\vec{r}) \right|_{\vec{r} \in \partial \Omega} = 0
\label{bcfn}
\end{eqnarray}

\begin{figure*}[t]
\hspace*{1.6cm} \includegraphics[width=0.9\textwidth]{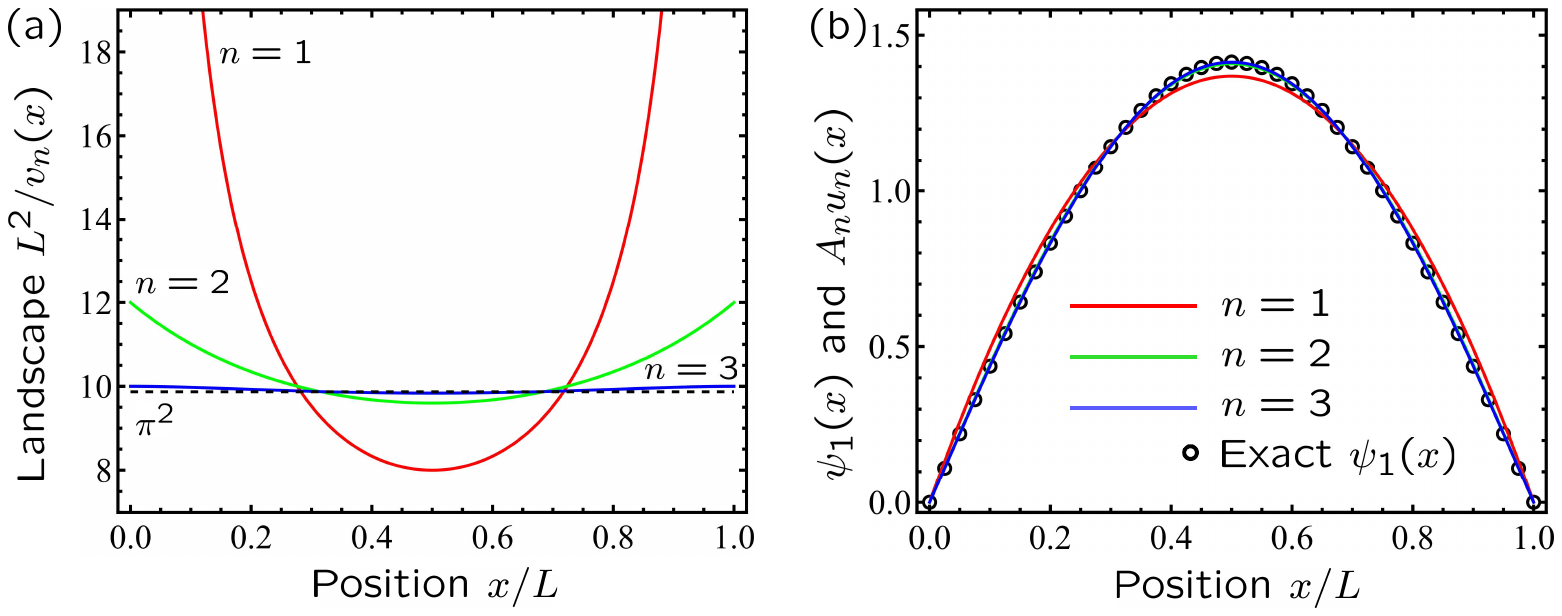}
\vspace*{-5.6cm}
\caption{\label{fig_box}
Higher-order landscape in a 1D box potential. (a) First three consecutive iterations for the landscape potential $1/v_n(x)$ and (b) the approximations $A_n u_n(x)$ of the fundamental wave function $\psi_1(x)$. Iterations rapidly converge to $1/v_{\infty}(x) = \pi^2/L^2 = E_1$ shown by the dashed line in (a) and $A_{\infty} u_{\infty}(x) = \psi_1(x)$ shown by open circles in (b).}
\end{figure*}

To summarize the proposed approach, we note that Eqs.\ (\ref{equnfinal}) and (\ref{eqphinfinal}) are exact at any order $n$. The wave function at energy $E$ can be reconstructed from their solutions as $\psi(\vec{r}) = u_n(\vec{r})\phi_n(\vec{r})$, with a subsequent normalization.
At any order $n$ we can choose to neglect the effective kinetic energy term $-\bm{\nabla} \cdot [u_n^2 \bm{\nabla} \phi_n]/u_n^2$ in Eq.\ (\ref{eqphinfinal}) and attempt to extract useful information from the landscape potential $1/v_n(\vec{r}) = u_{n-1}(\vec{r})/u_n(\vec{r})$.
$n=1$ yields the original localization landscape theory \cite{filoche12, arnold16} whereas $n = 2$ has been explored in Ref.\ \cite{chaudhuri20} and termed ``modified localization landscape theory''. The higher-order localization landscape theory presented in this work encompasses both these cases and goes beyond $n = 2$. 

When the kinetic energy term is neglected in Eq.\ (\ref{eqphinfinal}), the latter admits an exact solution if $v_n(\vec{r}) = u_n(\vec{r})/u_{n-1}(\vec{r})$  is independent of $\vec{r}$ and equal to $1/E$.  In this case, $u_{n-1}(\vec{r}) = E u_n(\vec{r})$, Eq.\ (\ref{equnfinal}) coincides with the Schrödinger equation (\ref{sch}) and thus
\begin{eqnarray}
\psi(\vec{r}) = A_n u_n(\vec{r})
\label{psiu}
\end{eqnarray}
with a normalization constant
\begin{eqnarray}
A_n = \left[ \int_{\Omega} d^d \vec{r}\, \left| u_n(\vec{r}) \right|^2 \right]^{-1/2}
\label{norma}
\end{eqnarray}
where $d^d \vec{r}$ denotes the volume element in a space of dimensionality $d$.
As we show in Appendix \ref{appproof},
these results become exact in the limit $n \to \infty$
with $E=E_1$ and $\psi(\vec{r}) = \psi_1(\vec{r})$. They can be seen as following from a known method in linear algebra (the so-called inverse power method \cite{bronson14}) intended to yield the smallest eigenvalue of a matrix (the matrix representing ${\hat H}$ in the configuration space) and the corresponding eigenvector. 

A similar reasoning can be applied if  $v_n(\vec{r})$ is piecewise constant. If the spatial domain $\Omega$ can be split in $M$ subdomains $\Omega_m$ in which $v_n(\vec{r}) \simeq v_n^{(m)}$ is approximately constant, Eqs.\ (\ref{equnfinal}) and (\ref{eqphinfinal}) allow for finding good approximations for energies $E_m = 1/v_n^{(m)}$ of locally fundamental eigenstates of Eq.\ (\ref{sch}) and for their wave functions (see Appendix \ref{appproof}):
\begin{eqnarray}
\psi_m(\vec{r}) \simeq A_n^{(m)} u_n(\vec{r}),\; \vec{r} \in \Omega_m
\label{psium}
\end{eqnarray}
with normalization constants
\begin{eqnarray}
A_n^{(m)} = \left[ \int_{\Omega_m} d^d \vec{r}\, \left| u_n(\vec{r}) \right|^2 \right]^{-1/2}
\label{normam}
\end{eqnarray}

The locally fundamental eigenstate of a subdomain $\Omega_m$ is the lowest-energy state among all states localized within $\Omega_m$ and having no nodes.
As we will see from the following, $v_n(\vec{r})$ becomes approximately piecewise constant for large but finite $n$, which is the main situation of interest here. 

\section{Examples of applications}
\label{examples}

In this section, we explore how the iterative higher-order landscape theory introduced above can be applied to analyze a number of simple problems.

\subsection{Box potential}
\label{box1d}

As the first and simplest example of application of the iterative scheme introduced in Sec.\ \ref{iter}, we consider a one-dimensional (1D) box potential: an interval $x \in [0,L]$ surrounded by infinitely high potential walls. This situation is described by $V(x) = 0$ and boundary conditions (\ref{bc}) at $x = 0, L$. The eigenvalues $E_m$ and eigenfunctions $\psi_m(x)$ of the Schr\"{o}dinger equation (\ref{sch}) with boundary conditions (\ref{bc}) are readily found \cite{landau}:
\begin{eqnarray}
E_m &=& \left( \frac{m \pi}{L} \right)^2
\label{enbox}
\\
\psi_m(x) &=& \sqrt{\frac{2}{L}} \sin\left( \frac{m \pi x}{L} \right)
\label{psinbox}
\end{eqnarray}
Applying the iterative scheme of Sec.\ \ref{iter} yields
\begin{eqnarray}
&&u_1(x) = \frac{x L}{2} \left( 1 - \frac{x}{L} \right)
\label{u1box}
\\
&&u_2(x) =\frac{x L^3}{24} \left [ 1 - 2 \left(\frac{x}{L}\right)^2 + \left(\frac{x}{L}\right)^3 \right]
\label{u2box}
\\
&&u_3(x) = \frac{x L^5}{240} \left [ 1 - \frac53 \left(\frac{x}{L}\right)^2 + \left(\frac{x}{L}\right)^4 - \frac13 \left(\frac{x}{L}\right)^5 \right] \;\;
\label{u3box}
\\
&&\ldots \nonumber
\\
&&v_1(x) = \frac{x L}{2} \left( 1 - \frac{x}{L} \right)
\label{v1box}
\\
&&v_2(x) =\frac{L^2}{12} \frac{1 - 2 \left(\frac{x}{L}\right)^2 + \left(\frac{x}{L}\right)^3}{1 - \frac{x}{L}}
\label{v2box}
\\
&&v_3(x) = \frac{L^2}{10} \frac{1 - \frac53 \left(\frac{x}{L}\right)^2 + \left(\frac{x}{L}\right)^4 - \frac13 \left(\frac{x}{L}\right)^5}{1 - 2 \left(\frac{x}{L}\right)^2 + \left(\frac{x}{L}\right)^3}
\label{v3box}
\\
&&\ldots \nonumber
\\
&&v_{\infty}(x) = \frac{L^2}{\pi^2}
\label{vinfbox}
\end{eqnarray}
Equations (\ref{u1box})--(\ref{vinfbox}) are illustrated in Fig.\ \ref{fig_box}. We see that $1/v_n(x)$ rapidly tends to $1/v_{\infty}(x) = \pi^2/L^2$ and $A_n u_n(x)$ converges to $\psi_1(x)$ as $n$ increases (see Appendix \ref{appproof} for a proof). In particular $A_3 u_3(x)$ is indistinguishable from $\psi_1(x)$ by eye in Fig.\ \ref{fig_box}(b) and the difference between the two does not exceed 0.15\% at any $x$. 

Let us now check whether the converged iterative solution $v_{\infty}(x)$ yields more accurate results than the first iteration $v_1(x) = u_1(x)$ corresponding to the standard landscape theory.
First, $1/\max[v_{\infty}(x)] = 1/v_{\infty}(x) = \pi^2/L^2$ is the exact value of the lowest eigenenergy $E_1$, in contrast to $1/\max[u_1(x)] = 8/L^2$ or the approximation $E_1 \simeq (1 + d/4)/\max[u_1(x)] = 10/L^2$ proposed earlier \cite{arnold19}. Next, if we follow Ref.\ \cite{david21} and approximate the integrated density of states (IDOS) ${\cal N}_n(E)$ by the (integer) number of subintervals of length $\pi/\sqrt{E}$ where the maximum of $v_n(x)$ is larger than $1/E$, we obtain
\begin{eqnarray}
{\cal N}_{\infty}(E) &=& \text{int} \left(
\frac{L}{\pi/\sqrt{E}}
\right) \Theta\left(\max\left[v_{\infty}(x) \right] - \frac{1}{E} \right)
\nonumber \\
&=& 
\text{int} \left(\frac{L}{\pi} \sqrt{E} \right) \Theta\left(E -\frac{1}{\max\left[v_{\infty}(x) \right]} \right)
\nonumber \\
&=& 
\text{int} \left(\frac{L}{\pi} \sqrt{E} \right) \Theta\left(E -\frac{\pi^2}{L^2} \right)
\label{idosbox}
\end{eqnarray} 
where $\text{int}(x)$ denotes the integer part of $x$ and $\Theta(x)$ is the Heaviside step function. 
It is easy to check that this expression coincides with the exact result for IDOS: ${\cal{N}}(E)$ increases by unity each time when $E$ reaches the energy $E_m = (m\pi/L)^2$ of a new eigenstate, starting from $m = 1$.

To summarize this subsection, we observe that our higher-order landscape theory allows us to obtain exact results for the first eigenenergy $E_1$ and IDOS in a 1D box potential.

\subsection{Pieces model}
\label{pieces}

Comtet and Texier have argued \cite{comtet20} that the (first-order) landscape theory fails to provide adequate results for
\begin{eqnarray}
V(x) &=& \sum\limits_n V_n \delta(x-x_n)
\label{piecesv}
\end{eqnarray} 
where $x_n$ are independently and uniformly distributed in $\Omega = [0,L]$ with mean density $\rho$, in the limit $V_n \to \infty$. In particular, the asymptotics of IDOS ${\cal N}_1(E) \simeq (L\sqrt{E}/2) \exp(-\rho \sqrt{8/E})$ does not match the expected ${\cal N}(E) \simeq \rho L \exp(-\pi \rho/\sqrt{E})$. A better approximation for ${\cal N}_1(E)$ proposed by Filoche {\it et al.} in response to this observation \cite{filoche20} still does not agree with the exact ${\cal N}(E)$.

Let us use our higher-order landscape theory to compute IDOS ${\cal N}(E)$ for the model (\ref{piecesv}). To this end, we note that ${\cal N}(E)$ is a superposition of IDOS of infinite square potential wells of width $\ell_n = x_n - x_{n-1}$ given by Eq.\ (\ref{idosbox}) with $L$ replaced by $\ell_n$. Averaging this equation over $\ell_n$ with a distribution $p(\ell) = \rho \exp(-\rho \ell)$ following from the independence of $x_n$ and multiplying the result by the number of wells $\rho L$, we obtain
\begin{eqnarray}
{\cal N}_{\infty}(E) &=& \rho L
\int\limits_0^{\infty} d\ell \rho e^{-\rho \ell}
\text{int} \left(\frac{\ell}{\pi} \sqrt{E} \right) \Theta\left(E -\frac{\pi^2}{\ell^2} \right)
\nonumber \\
&=&
\rho^2 L \frac{\pi}{\sqrt{E}}
\int\limits_0^{\infty} dt  e^{-\rho \pi t/\sqrt{E}}
\text{int} \left(t\right) \Theta\left(t-1\right)
\nonumber \\
&=&
\rho^2 L \frac{\pi}{\sqrt{E}}
\sum\limits_{n=1}^{\infty} n
\int\limits_n^{n+1} dt  e^{-\rho \pi t/\sqrt{E}}
\nonumber \\
&=&
\frac{\rho L}{\exp(\pi \rho/\sqrt{E}) -1}
\label{idospieces}
\end{eqnarray}
where we introduced $t = \ell \sqrt{E}/\pi$ to perform the integration. Equation (\ref{idospieces}) is actually the exact result for ${\cal N}(E)$ \cite{bychkov66,luttinger73}. Thus, our higher-order localization landscape theory allows us to obtain an exact analytical result for this particular problem and thus fixes the deficiency of the original, first-order landscape theory brought up in Ref.\ \cite{comtet20}.

\begin{figure*}[t]
\includegraphics[width=0.9\textwidth]{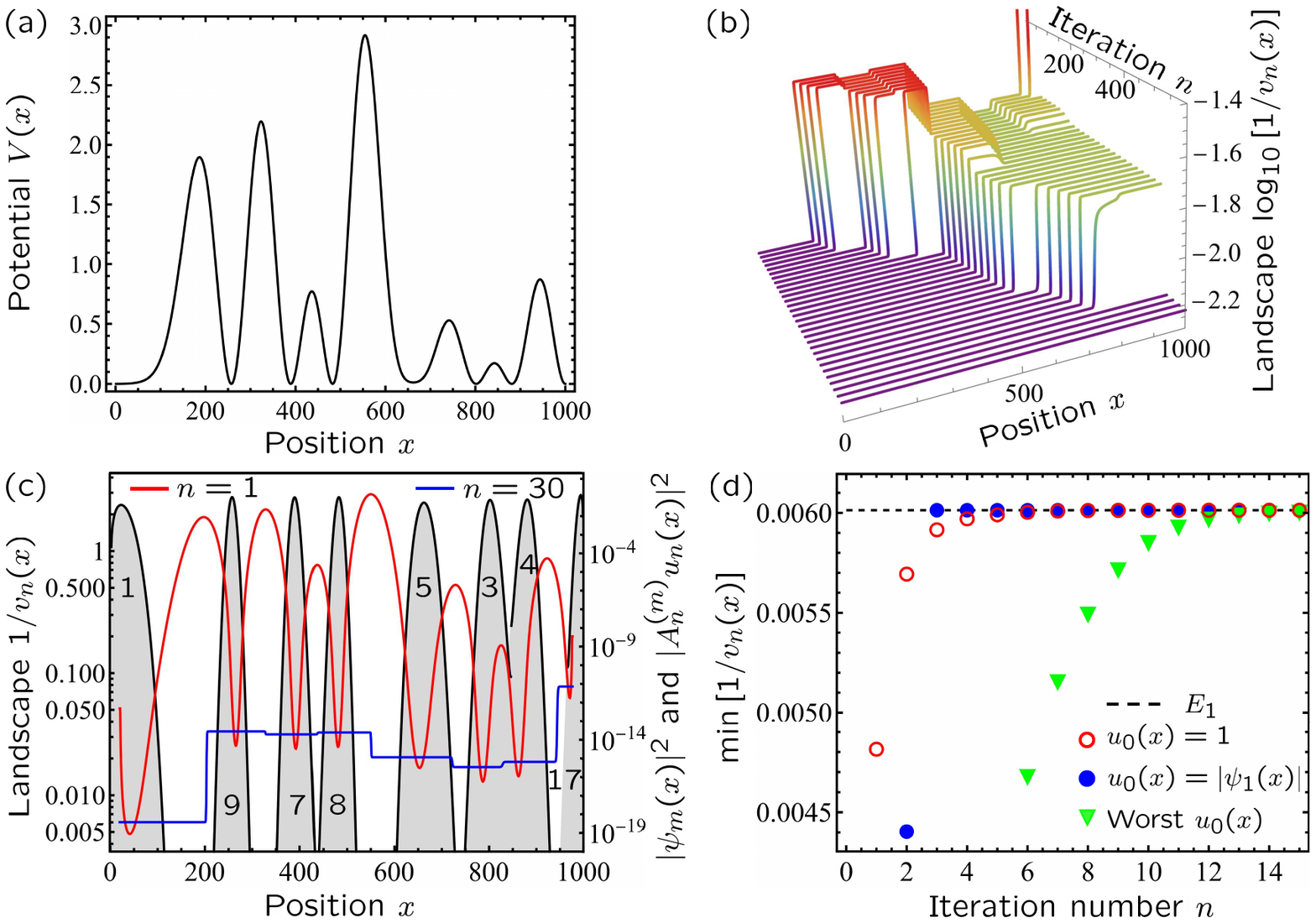}
\caption{\label{fig_smooth}
Higher-order landscape in a smooth potential. (a) A random potential obtained from Eq.\ (\ref{vcorr}) with $W = 1$ and (b) the associated higher-order landscape potentials for $n = 20$--600. Color scale highlights the magnitude of $1/v_n(x)$, with the violet for the lowest value and the red for the largest one.
(c) First- (red line) and thirtieth-order (blue line) landscape potentials (left scale). Gray shaded areas show intensities of wave functions $\psi_m(x)$ of locally fundamental eigenstates (right scale). Numbers are serial numbers $m$ of the corresponding eigenvalues $E_m$, with $E_m$ ordered in ascending order. Black lines are approximations (\ref{psium}) to the wave function intensities
{for $n = 30$}. (d) Predictions of the higher-order landscape theory (open circles) for the smallest eigenvalue $E_1$ in the random potential of panel (a) compared to the exact result (dashed line).
{Full blue circles and green triangles show the results obtained with $u_0(x) = |\psi_1(x)|$ and $u_0(x) = \Theta(x-L/2)$ (``worst case'') instead of $u_0(x) = 1$, respectively.} 
}
\end{figure*}

\subsection{Smooth random potential}
\label{smooth}

To simulate a smooth random potential on a 1D spatial domain $\Omega = [0, L]$, we consider a squared superposition of harmonic functions obeying zero boundary conditions at the boundaries $x = 0$ and $L$ of the domain and having random amplitudes:
\begin{eqnarray}
V(x) = W \left[
\sum\limits_{k=1}^{k_{\text{max}}} A_k \sin\left( \frac{\pi k x}{L} \right)
\right]^2
\label{vcorr}
\end{eqnarray}
where random $A_k$ are independent and uniformly distributed between $-1/2$ and $1/2$, and $W$ measures the strength of the potential. Figure \ref{fig_smooth}(a) shows a random realization of such a potential for $k_{\text{max}} = 10$.

\begin{table}
\begin{tabular}{c|c|c|c|c}
$m$  & $E_m$ & $\frac{1}{v_1\left[x_{\text{max}}^{(m)} \right]}$ & $\frac{1}{v_1\left[x_{\text{max}}^{(m)} \right]} \left(1 + \frac{d}{4} \right)$ & $\frac{1}{v_{30}\left[x_{\text{max}}^{(m)} \right]}$\\
\hline
1 & 0.0060131 & 0.0048140 & 0.0060175 & 0.0060131\\
3 & 0.0169974 & 0.0129417 & 0.0161771 & 0.0169974\\
4 & 0.0187773 & 0.0143156 & 0.0178945 & 0.0187773\\
5 & 0.0204678 & 0.0167233 & 0.0209041 & 0.0204589\\
7 & 0.0315125 & 0.0240505 & 0.0300631 & 0.0315125\\
8 & 0.0326643 & 0.0249348 & 0.0311685 & 0.0326643\\
9 & 0.0333359 & 0.0254707 & 0.0318384 & 0.0333359\\
17 & 0.0776201 & 0.0627534 & 0.0784418 & 0.0776201
\end{tabular}
\caption{\label{tab1} Eigenenergies $E_m$ of the eigenfunctions shown in Fig.\ \ref{fig_smooth}(c) (second column) compared to bare (third column) and refined (fourth column) predictions of the first-order landscape theory as well as to the prediction of the thirtieth-order landscape (last column).}
\end{table}

\begin{figure*}[t]
\includegraphics[width=0.9\textwidth]{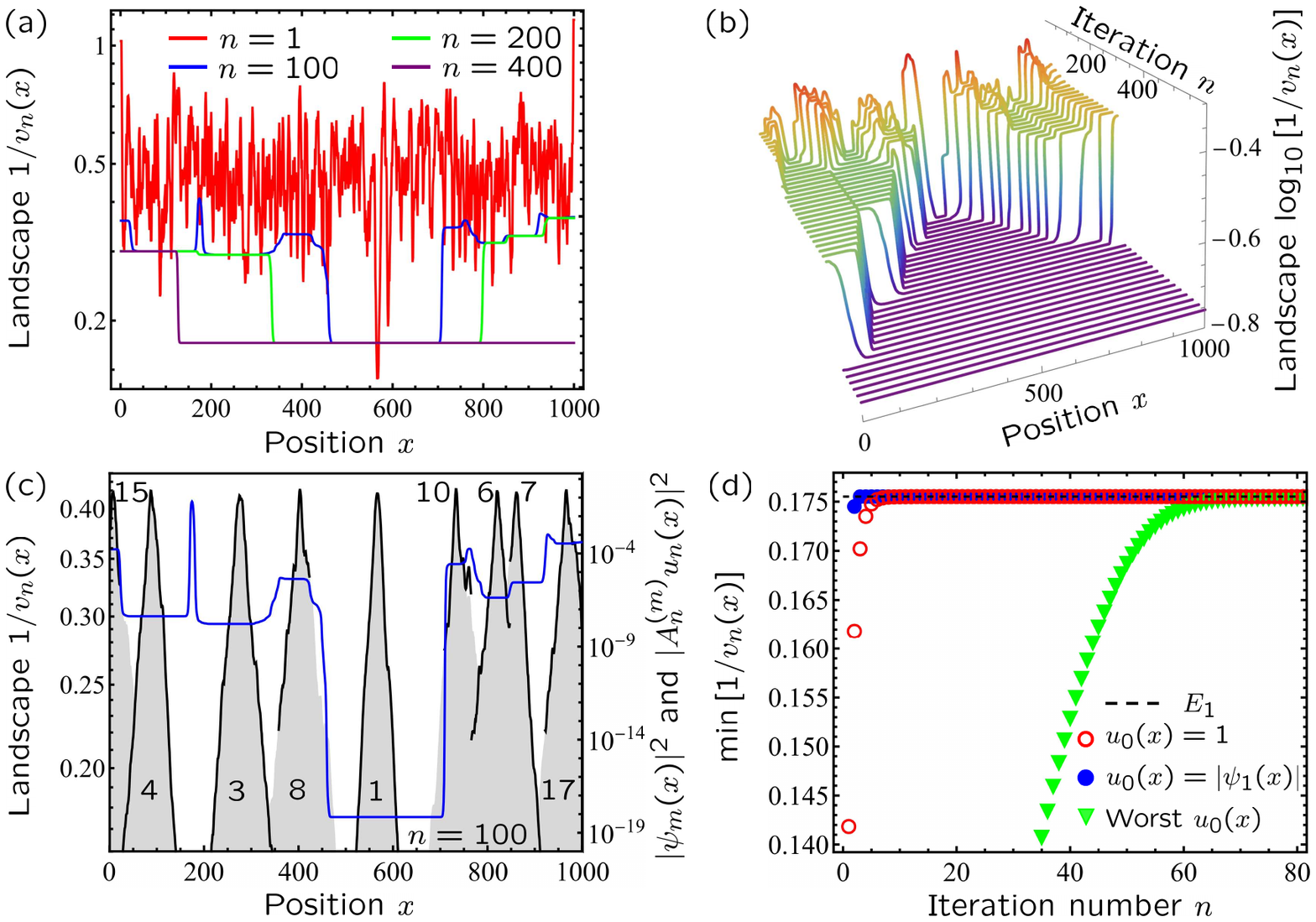}
\caption{\label{fig_white}
Higher-order landscape in an uncorrelated potential. (a) Landscape potentials of orders $n = 1$ (red), 100 (blue), 200 (green) and 400 (purple) for an uncorrelated (white-noise) random potential of amplitude $W=1$.
{(b) The same as (a) for $n=20$--600. Color scale highlights the magnitude of $1/v_n(x)$, with the violet for the lowest value and the red for the largest one.}
(c) Spatial profiles of several low-order eigenfunctions $\psi_m(x)$ (gray shaded areas, right scale) compared with the predictions (\ref{psium}) of the higher-order landscape theory ($n = 100$, black lines). Serial numbers $m$ of eigenfunctions are indicated by numbers. The corresponding landscape potential is shown by the blue line (left scale).  
(d) Predictions of the higher-order landscape theory (open circles) for the smallest eigenvalue $E_1$ compared to the exact result (dashed line).
{Full blue circles and green triangles show the results of using $u_0(x) = |\psi_1(x)|$ and $u_0(x) = 1 - \Pi(2x/L-1.1)$ (worst case) instead of $u_0(x) = 1$, respectively.} 
}
\end{figure*}

We set $L = 10^3$, discretize the domain $\Omega$ with a discretization step $\Delta x = 1$, and solve both the eigenvalue problem (\ref{sch}) and landscape equations (\ref{un})--(\ref{bcfn}) numerically, see Appendix \ref{appnum}. As the order $n$ of iteration increases, $v_n(x)$ rapidly converges to a roughly piecewise constant function.
Further iterations slowly lead to merging of domains in which $v_n(x)$ is constant and progressive extension of the domain in which $v_n(x)$ is the largest (i.e. $1/v_n(x)$ is the smallest),
{see Fig.\ \ref{fig_smooth}(b).}
The latter extends to the whole $\Omega$ for large enough $n$, so that $v_{\infty}(x) = 1/E_1$ and $A_{\infty} u_{\infty}(x) = \psi_1(x)$ yield exact results for the energy $E_1$ and the wave function $\psi_1(x)$ of the fundamental eigenstate of the original problem (\ref{sch}). Useful information about other low-energy eigenstates can be obtained from intermediate $n$ for which $v_n(x)$ has already acquired an approximately piecewise constant form but has not converged to a position-independent constant yet. Figure \ref{fig_smooth}(c) presents a comparison of the first-order landscape potential ($n = 1$, red line) with the thirtieth-order one ($n = 30$, blue line). It also shows intensities of wave functions of eight locally fundamental eigenstates of subdomains $\Omega_m$ (gray shading) compared to their approximations by Eq.\ (\ref{psium}) (black lines) that turn out to be excellent.

Values of $1/v_n^{(m)}$ inside domains $\Omega_m$ of approximate constancy of $v_n(x)$ provide good approximations to the energies $E_m$ of locally fundamental eigenstates of $\Omega_m$. If we define $x_{\text{max}}^{(m)}$ as the position of the $m$-th local maximum of $u_n(x)$:
\begin{eqnarray}
\max_{\Omega_m} u_n(x) = u_n\left[ x_{\text{max}}^{(m)} \right]
\label{xmaxdef}
\end{eqnarray}
then $1/v_n[x_{\text{max}}^{(m)}]$ yields $E_m$ with great accuracy (compare the second and last columns in Table\ \ref{tab1}). For comparison, Table\ \ref{tab1} also shows approximations to $E_m$ obtained from the original, first-order landscape theory (third column) as well as from its refined version (fourth column).
A good idea about the convergence of $1/v_n[x_{\text{max}}^{(m)}]$ to $E_m$ can be obtained by looking at the accuracy of approximation of the lowest eigenenergy $E_1$ by $\min[1/v_n(x)]$. A comparison between $\min[1/v_n(x)]$ (red open circles)
and $E_1$ in Fig.\ \ref{fig_smooth}(d) demonstrates very good agreement between the two already for $n = 10$ and even below.

{Note that the simple relations between $u_n(x)$, $v_n(x)$ and $\psi_m(x)$, $E_m$ demonstrated above hold only for the locally fundamental eigenstates. Information about oscillating eigenfunctions corresponding to $m = 2$, 6, 10--16 and absent from Fig.\ \ref{fig_smooth}(c) and Table\ \ref{tab1}, cannot be easily obtained from $u_n(x)$ because these states provide only a minor contribution to $u_n(x)$, as discussed in Appendix\ \ref{appproof}. They may cause slight deviations of the predictions of the higher-order landscape theory from the exact results for the locally fundamental eigenstates with which they share spatial localization domains. This is the case for $m = 5$ in Table\ \ref{tab1} because oscillating states corresponding to $m = 10$ and 15 are localized in the same range of $x$.
} 

{
The iterative procedure defined by Eq.\ (\ref{equnfinal}) can be initiated with $u_0(\vec{r})$ different from 1. In principle, it may be possible to tailor $u_0(\vec{r})$ to optimize a desired aspect of the calculation. We test this possibility for the case of the smooth potential of Fig.\ \ref{fig_smooth}(a) by studying the impact of $u_0(x)$ on the convergence of $\min[1/v_n(x)]$ to $E_1$. On the one hand, we find that tested choices of $u_0(x) \ne 1$ that do not imply knowing any information about the eigenfunctions of ${\hat H}$ [such as, e.g., $u_0(x) = 1/V(x)$ or randomly fluctuating $u_0(x)$] do not speed up convergence compared to $u_0(x) = 1$. They may have a strong influence on the result of the first iteration but only weakly affect further convergence, almost always resulting in $\min[1/v_n(x)]$ being very close to $E_1$ for $n \simeq 10$. On the other hand, because we expect the spatial profile of $u_n(x)$ to coincide with $\psi_1(x)$ up to a numerical prefactor for large $n$ [see Eq.\ (\ref{psiu}) and the following discussion, as well as Appendix  \ref{appproof}], choosing $u_0(x)$ close to the fundamental eigenfunction improves convergence. This is illustrated by full blue circles in Fig.\ \ref{fig_smooth}(d) for $u_0(x) = |\psi_1(x)|$ that converge to $E_1$ after three iterations, even though the results of the first two iterations are further from $E_1$ than those for $u_0(x) = 1$. In contrast, if $u_0(x)$ has little overlap with $\psi_1(x)$, convergence may slow down significantly, as we see in Fig.\ \ref{fig_smooth}(d) for $u_0(x) = \Theta(x-L/2)$ (green triangles). Such a choice of $u_0(x)$ is disadvantageous and probably one of the worst because $\psi_1(x)$ is localized near $x=0$ [see  Fig.\ \ref{fig_smooth}(c)]. However, even such a bad choice of $u_0(x)$ does not preclude convergence, testifying that the convergence of $\min[1/v_n(\vec{r})]$ to $E_1$ is quite robust with respect to the choice of $u_0(\vec{r})$. It remains to be seen to what extent the choice of $u_0(\vec{r})$ affects other aspects of the presented theoretical approach, such as, e.g., the structure of plateaus in Fig.\ \ref{fig_smooth}(b).
}

\subsection{Uncorrelated random potential}
\label{white}

A paradigmatic model of a disordered quantum system is the Anderson tight-binding model with an uncorrelated random potential. We obtain it by assuming independent random values of the potential $V$
{in the range $[0,W]$}
at discrete lattice points $x_k = k \Delta x$ of the 1D spatial grid introduced in the previous subsection.
Solving discretized versions of higher-order landscape equations (\ref{un})--(\ref{bcfn})
{(see Appendix \ref{appnum})}
yields landscape potentials shown in Figs.\ \ref{fig_white}(a) and (b). They present a behavior similar to that for the smooth potential (see Sec.\ \ref{smooth}): for large enough $n$, $v_n(x)$ is roughly a piecewise constant function, with the number of constancy domains $\Omega_m$ decreasing with $n$ and merging into a single domain $\Omega$ in which $v_n(x) = 1/E_1$ in the limit of $n \to \infty$. Again, Eq.\ (\ref{psium}) provides very good approximations for locally fundamental eigenfunctions as we illustrate in Fig.\ \ref{fig_white}(c) and the minimum of $1/v_n(x)$ converges rapidly to the fundamental eigenenergy $E_1$, see Fig.\ \ref{fig_white}(d).
{
Similarly to the case of smooth potential in Sec.\ \ref{smooth}, choosing $u_0(x)$ different from 1 impacts the convergence, with $u_0(x) = |\psi_1(x)|$ ensuring a faster convergence, see the full blue circles in Fig.\ \ref{fig_white}(d). As a worst-case scenario, we consider $u_0(x) = 1 - \Pi(2x/L-1.1)$ that is zero for $300 < x < 800$ where the first eigenfunction is localized [see Fig.\ \ref{fig_white}(c)] and one elsewhere. Here $\Pi(x)$ is the rectangular (box) function. Such a bad choice of $u_0(x)$ considerably slows down the convergence, requiring around 70 iterations for $\min[1/v_n(x)]$ to reach $E_1$, but still does not preclude it. 
}  
 
Energies of locally fundamental eigenstates of domains $\Omega_m$ are equal to $1/v_n[x_{\text{max}}^{(m)}]$ with a very good accuracy, see Table \ref{tab2}. Slight deviations of  the predictions of the higher-order landscape theory from exact results arise for states $m = 7$ and 17
{
and can be attributed to states with oscillating eigenfunctions localized in the same spatial domains, see the discussion in Sec.\ \ref{smooth}.}

\begin{figure}[t]
\includegraphics[width=0.48\textwidth]{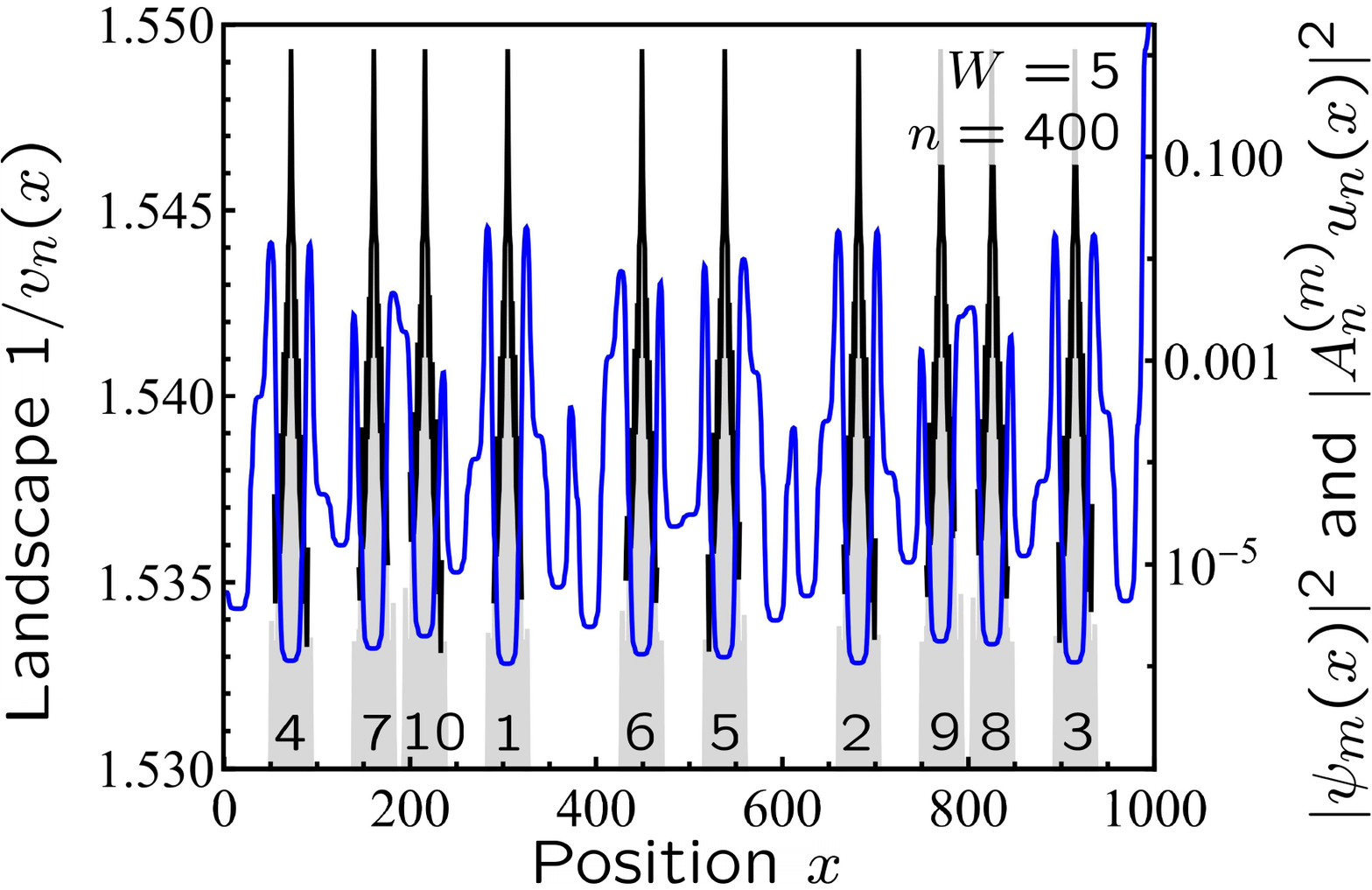}
\caption{\label{fig_aa}
Higher-order landscape in a quasiperiodic potential. Same as Figs.\ \ref{fig_smooth}(c) and \ref{fig_white}(c) but for the Aubry-Andr\'{e} model with $W = 5$.
}
\end{figure}

\subsection{Quasiperiodic potential}
\label{quasi}

We use the same discrete lattice as in the two previous subsections and consider the Aubry-Andr\'{e} model \cite{harper55,aubry80}:
\begin{eqnarray}
V_k=  \frac{W}{2} \left[ 1 + \cos(2\pi\gamma k + \varphi)
\right]
\label{aa}
\end{eqnarray}
with an irrational $\gamma = (\sqrt{5} - 1)/2$ and $\varphi = 0$. An interesting aspect of this model is that it exhibits a localization transition at $W = W_c = 4$: all states are extended for $W < W_c$ whereas spatially localized states arise for $W > W_c$ \cite{boers07,modugno09,li17aa}. Figure\ \ref{fig_aa} shows the localization landscape of a relatively high order $n=400$ (blue line) together with the ten first low-energy eigenstates $\psi_m(x)$ (gray shading) and their approximations by Eq.\ (\ref{psium}) (black lines) for $W = 5 > W_c$. Again, the approximation provided by Eq.\ (\ref{psium}) is excellent. The agreement between the lowest $E_m$ and their estimations $1/v_n[x_{\text{max}}^{(m)}]$ (not shown) is also good, similarly to what we show in Tables \ref{tab1} and \ref{tab2} for the models studied in Secs.\ \ref{smooth} and \ref{white}. It is interesting to note that the quasiperiodicity of the physical potential $V(x)$ leads to the quasiperiodicity of the landscape potential $1/v_n(x)$ that clearly exhibits some structure.   

\begin{table}
\begin{tabular}{c|c|c|c|c}
$m$  & $E_m$ & $\frac{1}{v_1\left[x_{\text{max}}^{(m)} \right]}$ & $\frac{1}{v_1\left[x_{\text{max}}^{(m)} \right]} \left(1 + \frac{d}{4} \right)$ & $\frac{1}{v_{100}\left[x_{\text{max}}^{(m)} \right]}$\\
\hline
1 & 0.175509 & 0.143133 & 0.178916 & 0.175509\\
3 & 0.294099 & 0.248550 & 0.310687 & 0.294099\\
4 & 0.300183 & 0.230402 & 0.288002 & 0.300183\\
6 & 0.315470 & 0.244650 & 0.305813 & 0.315470\\
7 & 0.328613 & 0.263166 & 0.328958 & 0.328608\\
8 & 0.331643 & 0.250712 & 0.313390 & 0.331643\\
10 & 0.345092 & 0.268219 & 0.335274 & 0.345092\\
15 & 0.359089 & 0.301749 & 0.377186 & 0.359089\\
17 & 0.364635 & 0.277763 & 0.347204 & 0.364631\\
\end{tabular}
\caption{\label{tab2} Eigenenergies $E_m$ of the eigenfunctions shown in Fig.\ \ref{fig_white}(c) (second column) compared to bare (third column) and refined (fourth column) predictions of the first-order landscape theory as well as to the prediction of the higher-order landscape theory of order $n = 100$ (last column).}
\end{table}

\begin{figure*}[t]
\hspace*{1.6cm} \includegraphics[width=\textwidth]{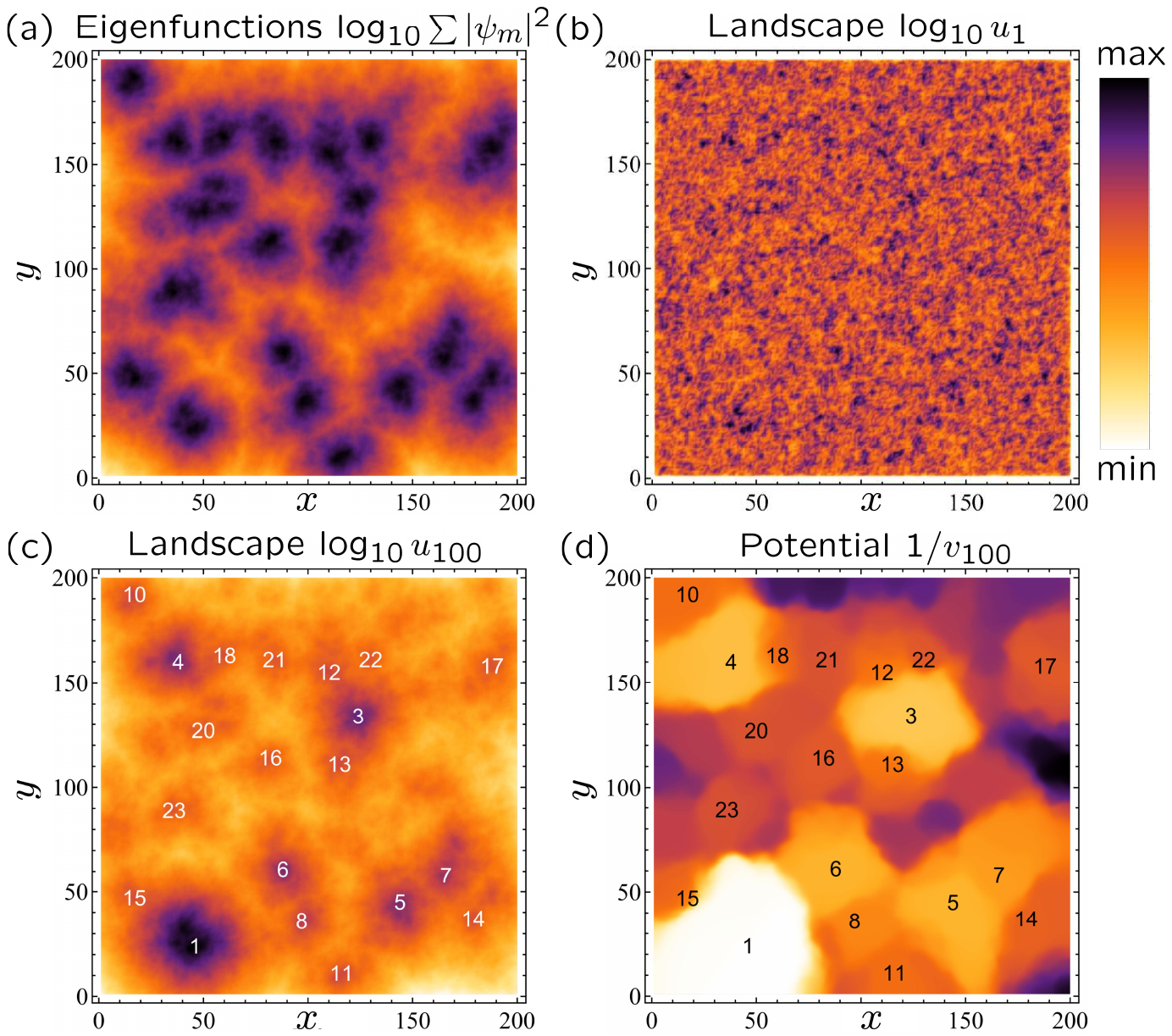}
\caption{\label{fig_2d}
{Higher-order landscape in 2D. (a) Sum of absolute values squared of the first 20 locally fundamental eigenfunctions in a 2D uncorrelated (white-noise) random potential of amplitude $W = 2$ on a $200 \times 200$ lattice.  (b) First-order localization landscape function $u_1(\vec{r})$. (c) Hundredth-order localization landscape function $u_{100}(\vec{r})$. (d) Hundredth-order effective landscape potential $1/v_{100}(\vec{r})$. Numbers $m = 1$--23 in (c) and (d) indicate positions of the maxima of eigenfunctions in (a) ordered in the order of increasing energy $E_m$. The color scale is adjusted to cover the full range of values for each panel separately. Equations for $u_n$ are solved numerically by adapting the method described in Appendix \ref{appnum} to 2D.}}
\end{figure*}

\subsection{Higher dimensions}
\label{highd}

{
Even though all the examples considered up to now were 1D, the formulation of the higher-order landscape theory given in this work applies in arbitrary dimension $d$. Advantages of using this approach in dimensions $d > 1$ remain to be explored, but we nevertheless provide here an example of its application for the uncorrelated random potential in 2D analogous to that considered in Sec.\ \ref{white}. Figure\ \ref{fig_2d}(a) shows spatial profiles of the first 20 locally fundamental eigenfunctions obtained for a single random realization of such a potential of strength $W = 2$. Note that the eigenfunctions corresponding to $m =2$, 9 and 19 are not locally fundamental and are not shown. Their spatial locations coincide with one of the locally fundamental modes. Computing the standard (first-order) localization landscape $u_1$ yields a function shown in Fig.\ \ref{fig_2d}(b). Its close inspection allows for identifying maxima at positions of eigenfunctions in (a) but the structure of $u_1(\vec{r})$ is obviously much richer than that resulting from 20 lowest-order eigenfunctions in Fig.\ \ref{fig_2d}(a). Calculating a higher-order landscape $u_n(\vec{r})$ allows for reducing the amount of information contained in the landscape function and making the latter a better predictor of the spatial profiles of the lowest-order eigenfunctions as we illustrate in Fig.\ \ref{fig_2d}(c) for $n = 100$. Similarly to 1D, the landscape potential $1/v_{n}(\vec{r})$ is composed of flat regions $\Omega_m$ and looks like a patchwork, see Fig.\ \ref{fig_2d}(d). Values of $1/v_{n}^{(m)}$ in each $\Omega_m$ provide very good approximations to eigenenergies $E_m$ of locally fundamental eigenfunctions $\psi_m(\vec{r})$.
}

\section{Conclusions}
\label{concl}

In this paper, we introduce an extension of the localization landscape theory by Filoche and Mayboroda \cite{filoche12} which amounts to solving an equation ${\hat H}^n u_n(\vec{r}) = 1$ for increasing integer orders $n$. The landscape potential $1/v_n(\vec{r}) = u_{n-1}(\vec{r})/u_{n}(\vec{r})$ tends to acquire a piecewise constant form $1/v_n(\vec{r}) \simeq 1/v_n^{(m)}$ in spatial domains $\vec{r} \in \Omega_m$ as $n$ increases. Eigenenergies of locally fundamental eigenstates can be estimated as $E_m = 1/v_n^{(m)}$ whereas their wave functions $\psi_m(\vec{r})$ can be approximated by $u_n(\vec{r})$ up to a normalization. The number of domains $\Omega_m$ decreases with $n$, reducing the number of $E_m$ and $\psi_m(\vec{r})$ that can be estimated at a given $n$ but increasing the accuracy of estimations.
{In practice, a good compromise is achieved for $n$ that ensures that the plateaus of $1/v_n(\vec{r})$ are well identified and sufficiently extended. The corresponding value of $n$ depends on the nature of the random potential and was found to be of the order of 10--100 for the examples considered in this work.}
The smallest eigenenergy $E_1$ and the corresponding eigenfunction $\psi_1(\vec{r})$ are obtained exactly in the limit $n \to \infty$, which corresponds to a known result of the inverse power method in linear algebra \cite{bronson14}. For simple systems---a box potential and a sequence of randomly located impenetrable delta-function potential barriers---where the original landscape theory already yields good but still approximate results, our approach provides exact analytic results for the energy of the lowest eigenstate (for the box potential) and the integrated density of states (IDOS, for both systems). Numerical solutions for random and quasiperiodic potentials allow for estimating the eigenenergies of several (up to 20 in the considered examples) locally fundamental eigenstates as well as for reconstructing their spatial structures with very high accuracy. A possible extension of the proposed theoretical framework concerns the problem of many-body localization \cite{bala20}. 

{It should be noted that the higher-order landscape potential is not necessarily superior to the standard, first-order one in all aspects. A representative example is the approximation to IDOS following from Weyl's law with the potential $V(x)$ replaced by the effective potential $1/v_n(x)$, by analogy with the proposal of Ref.\ \cite{arnold16} for $n = 1$. Our calculations for the uncorrelated potential in 1D show that a large-$n$ result can be very accurate in pinpointing the precise energy at which IDOS becomes different from zero but performs worse than the $n=1$ one in predicting IDOS at higher energies.
}

In addition to yielding accurate numerical results, the proposed theoretical framework suggests that the original landscape theory \cite{filoche12} can be seen as the first iteration of the inverse power method for finding the smallest eigenvalue of a matrix \cite{bronson14}. The modified localization landscape theory of Ref.\ \cite{chaudhuri20} corresponds to the second iteration. Without diminishing the merits of these theories, such a point of view may help to understand at least one of the reasons for their success without sophisticated mathematical proofs.


\acknowledgments
I acknowledge discussions with Marcel Filoche and the support from the Simons foundation that allowed me to participate in the 2024 Simons Collaboration on Localization of Waves Annual Meeting that largely motivated this work.

\appendix

\section{{Landscape functions for large $n$}}
\label{appproof}

The solution $u_n(\vec{r})$ of Eq.\ (\ref{equnfinal}) can be expanded in eigenfunctions $\psi_m(\vec{r})$ of the Schr\"{o}dinger equation (\ref{sch}):
\begin{equation}
u_n(\vec{r})  = \sum\limits_{m=1}^{\infty} a_{nm} \psi_m(\vec{r})
\label{expansion}
\end{equation}  
Substituting Eq.\ (\ref{expansion}) into Eq.\ (\ref{equnfinal}) yields
\begin{equation}
a_{nm} = \frac{a_{(n-1)m}}{E_m} = \frac{a_{0m}}{E_m^n}
\label{anm}
\end{equation}  
where
\begin{equation}
a_{0m} = \int_{\Omega} d^d \vec{r}\; \psi_m^*(\vec{r}) u_0(\vec{r})
\label{a0m}
\end{equation}  
and we have used ${\hat H} \psi_m = E_m \psi_m$. 
For $E_m$ ordered in ascending order and $a_{01} \ne 0$, we have
\begin{equation}
\begin{aligned}
\lim\limits_{n \to \infty} u_{n}(\vec{r}) &= a_{n1} \psi_1(\vec{r})
\\
\lim\limits_{n \to \infty} v_{n}(\vec{r}) &= \lim\limits_{n \to \infty} \frac{u_n(\vec{r})}{u_{n-1}(\vec{r})} = \frac{a_{n1}}{a_{(n-1)1}} = \frac{1}{E_1}
\end{aligned}
\label{uvconv}
\end{equation}
which coincides with Eq.\ (\ref{psiu}) for $\psi(\vec{r}) = \psi_1(\vec{r})$ and $A_n = 1/a_{n1}$.

According to Eqs.\ (\ref{expansion})--(\ref{a0m}), locally fundamental eigenstates $\psi_m(\vec{r})$ (i.e., the states that have no nodes) provide major contributions to $u_n(\vec{r})$ as far as $u_0(\vec{r}) = 1$ because states $\psi_m(\vec{r})$ that oscillate as a function of position $\vec{r}$ have much smaller $a_{0m}$. For this reason, the landscape functions $u_n(\vec{r})$ provide information about spatial profiles and eigenenergies of locally fundamentally eigenstates only. This property of higher-order landscape is inherited from the standard (first-order) landscape theory, for which it is discussed in Ref.\ \cite{filoche17}.

For $u_0(\vec{r}) = 1$ and the 1D box potential considered in Sec.\ \ref{box1d}, 
\begin{equation}
\begin{aligned}
\psi_m(x) &= \sqrt{\frac{2}{L}} \sin\left(\frac{m \pi x}{L} \right)\\
E_m &= \left( \frac{m \pi}{L} \right)^2\\
a_{0m} &= \frac{\sqrt{2L}}{m \pi} \left[
1 + (-1)^{m+1} \right]\\
a_{nm} &= \sqrt{\frac{2}{L}} \left( \frac{L}{m\pi} \right)^{2n + 1}
\left[ 1 + (-1)^{m+1} \right]\\
\lim\limits_{n \to \infty} u_{n}(x) &= \frac{4}{\pi} \left( \frac{L}{\pi} \right)^{2n}
\sin\left(\frac{\pi x}{L} \right)\\
\lim\limits_{n \to \infty} v_{n}(x) &= \frac{L^2}{\pi^2}
\end{aligned}
\label{uvconv}
\end{equation}
or, equivalently, 
\begin{eqnarray}
\begin{aligned}
\lim\limits_{n \to \infty} \frac{L^2}{v_{n}(x)} &= \pi^2
\end{aligned}
\label{vnconv2}
\end{eqnarray}
as illustrated in Fig.\ \ref{fig_box}(a).

If eigenfunctions $\psi_m(\vec{r})$ are spatially localized, then for a finite value of $n$, $u_n(\vec{r})$ in a subdomain $\Omega_m$ around the localization center $\vec{r}_m$ of the state $m$ can be approximated by a single term in the sum (\ref{expansion}):
\begin{equation}
u_n(\vec{r})  \simeq a_{nm} \psi_m(\vec{r}),\;\;
\vec{r} \in \Omega_m
\label{udomain}
\end{equation}  
Then,
\begin{equation}
v_n(\vec{r})  \simeq \frac{a_{nm}}{a_{(n-1)m}} = \frac{1}{E_m},\;\; \vec{r} \in \Omega_m
\label{vdomain}
\end{equation}  

\section{Numerical solution of the iterative landscape equation}
\label{appnum}

In order to solve Eq.\ (\ref{equnfinal}) in 1D, we introduce a fictitious time variable $t$ and consider a partial differential equation 
\begin{eqnarray}
\frac{\partial u_n(x,t)}{\partial t}  = \left[ \Delta - V(x) \right] u_n(x,t) + u_{n-1}(x,t)
\label{diffusion}
\end{eqnarray}
This equation can be discretized on a lattice $x_i = i \Delta x$: 
\begin{eqnarray}
\begin{aligned}
&\frac{u_n[i,t+\Delta t] - u_n[i,t]}{\Delta t} = 
\\
&\frac{u_n[i-1,t] - 2 u_n[i,t] + u_n[i+1,t]}{(\Delta x)^2}
\\
&- V[i] u_n[i,t] + u_{n-1}[i,t]
\end{aligned}
\label{discr1}
\end{eqnarray}
which is known as ``forward time centered space'' finite-difference approximation \cite{press92}.
Expressing $u_n[i,t+\Delta t]$ yields
\begin{eqnarray}
\begin{aligned}
&u_n[i,t+\Delta t]  = u_n[i,t]
\left(
1 - \frac{2 \Delta t}{(\Delta x)^2} - V[i] \Delta t
\right) u_n[i,t]
\\
&+ \frac{\Delta t}{(\Delta x)^2}
\left(
u_n[i-1,t] + u_n[i+1,t]
\right) + u_{n-1}[i,t] \Delta t
\end{aligned}
\label{discr2}
\end{eqnarray}
We choose $\Delta t/(\Delta x)^2 \leq 1/2$ to ensure numerical stability \cite{press92} and start with some initial values of $u_n[i,0]$ at $t = 0$ ($u_n[i,0] = 0$ in most cases). Equation\ (\ref{discr2}) is iterated until a sufficiently large time $t$ that ensures converges to a time-independent regime or, more precisely, until the absolute value of the modification of $u_n$ from one iteration to another becomes less than some small $\epsilon$ for all $x$ (typically, we set $\epsilon = 10^{-10}$).

\newpage
\bibliographystyle{apsrev4-2}
\bibliography{landscaperefs}

\end{document}